\documentclass[reprint,superscriptaddress,nofootinbib,amsmath,amssymb,aps,prl]{revtex4-1}

\usepackage{color}
\usepackage{braket}
\usepackage{graphicx}
\usepackage{dcolumn}
\usepackage{bm}
\usepackage[mathlines]{lineno}

\def\Tr{\mbox{Tr}}

\begin{document}

\title{Non-equilibrium quantum-heat statistics under stochastic projective measurements}

\author{Stefano Gherardini}
\email{gherardini@lens.unifi.it}
\affiliation{\mbox{Department of Physics and Astronomy, University of Florence,} via G. Sansone 1, I-50019 Sesto Fiorentino, Italy.}
\affiliation{\mbox{LENS, CNR-INO, and QSTAR,} via N. Carrara 1, I-50019 Sesto Fiorentino, Italy.}
\affiliation{\mbox{INFN Sezione di Firenze}, via G. Sansone 1, I-50019 Sesto Fiorentino, Italy.}

\author{Lorenzo Buffoni}
\affiliation{\mbox{Department of Physics and Astronomy, University of Florence,} via G. Sansone 1, I-50019 Sesto Fiorentino, Italy.}
\affiliation{\mbox{Department of Information Engineering, University of Florence,} via S. Marta 3, I-50139 Florence, Italy.}

\author{Matthias M. M\"uller}
\affiliation{\mbox{Department of Physics and Astronomy, University of Florence,} via G. Sansone 1, I-50019 Sesto Fiorentino, Italy.}
\affiliation{\mbox{LENS and QSTAR, University of Florence,} via G. Sansone 1, I-50019 Sesto Fiorentino, Italy.}

\author{Filippo Caruso}
\affiliation{\mbox{Department of Physics and Astronomy, University of Florence,} via G. Sansone 1, I-50019 Sesto Fiorentino, Italy.}
\affiliation{\mbox{LENS and QSTAR, University of Florence,} via G. Sansone 1, I-50019 Sesto Fiorentino, Italy.}

\author{Michele Campisi}
\affiliation{\mbox{Department of Physics and Astronomy, University of Florence,} via G. Sansone 1, I-50019 Sesto Fiorentino, Italy.}
\affiliation{\mbox{INFN Sezione di Firenze}, via G. Sansone 1, I-50019 Sesto Fiorentino, Italy.}

\author{Andrea Trombettoni}
\affiliation{\mbox{CNR-IOM DEMOCRITOS Simulation Center}, via Bonomea 265, I-34136 Trieste, Italy.}
\affiliation{\mbox{SISSA}, via Bonomea 265, I-34136 Trieste, Italy \& \mbox{INFN}, Sezione di Trieste, I-34151 Trieste, Italy.}

\author{Stefano Ruffo}
\affiliation{\mbox{SISSA}, via Bonomea 265, I-34136 Trieste, Italy \& \mbox{INFN}, Sezione di Trieste, I-34151 Trieste, Italy.}
\affiliation{Istituto  dei  Sistemi  Complessi, CNR, via  Madonna  del  Piano 10, I-50019 Sesto Fiorentino, Italy.}

\begin{abstract}
In this paper we aim at characterizing the effect of stochastic fluctuations on the distribution of the energy exchanged by a quantum system with an external environment under sequences of quantum measurements performed at random times. Both quenched and annealed averages are considered. The information about fluctuations is encoded in the quantum-heat probability density function, or equivalently in its characteristic function, whose general expression for a quantum system with arbitrary Hamiltonian is derived. We prove that, when a stochastic protocol of measurements is applied, the quantum Jarzynski equality is obeyed. Therefore, the fluctuation relation is robust against the presence of randomness in the times intervals between measurements. Then, for the paradigmatic case of a two-level system, we analytically characterize the quantum-heat transfer. Particular attention is devoted to the limit of large number of measurements and to the effects caused by the stochastic fluctuations. The relation with the stochastic Zeno regime is also discussed. \\
\begin{description}
\item[PACS numbers]
05.70.Ln, 03.65.Yz, 03.65.Ta, 05.40.-a
\end{description}
\end{abstract}

\maketitle

\section{\label{sec:intro}I. Introduction}

A growing interest in thermodynamic properties of quantum systems emerged in the last decade~\cite{Esposito2009,Campisi2011}. One of the main goals of such research activity is to devise and implement more efficient engines by exploiting quantum resources~\cite{Scully03Science299,Kosloff14ARPC65,Uzdin15PRX5,Campisi15NJP17}. In this direction,
it has been explored the role of non-thermal states~\cite{Solinas2015} and the capability of characterizing the statistics of the energy, which is exchanged by a quantum system in interaction with an external environment and/or measurement apparatuses~\cite{Campisi2010PRL,Campisi2011PRE,Yi2013,WatanabePRE2014,Talkner16PRE93,Yuanjian16PRE94}.

Quantum measurements, at variance with classical ones, are invasive and are accordingly accompanied by stochastic energy exchanges between the measurement apparatus and the measured system. In this work we adopt the convention to call such energy exchanges \emph{quantum-heat}~\cite{Elouard2016}, and denote it by the symbol $Q_q$, to distinguish it from the usual definition of the heat $Q$ (i.e., the energy exchanged with a thermal bath) and the work $W$ (i.e., the energy exchanged with a work source).

Randomness may appear in a measurement process, not only in the outcome of the measurement, but also in the time of its occurrence. Particularly interesting is the study of  stochastic measurement sequences consisting in repeated interactions between a quantum system and a measurement apparatus occurring at random
times~\cite{Gherardini2016NJP,Mueller2017ADP,GherardiniPhDThesis}. Such sequences have been studied both theoretically and experimentally in the context of Zeno dynamics on an atom-chip~\cite{SchaferZeno}, while their ergodicity properties have been investigated in~\cite{Gherardini2017QSc}.

In this paper, we study the statistics of the energy exchanged between a quantum system and a measurement apparatus, under the assumption that the interaction can be modelled  by a sequence of projective measurements occurring instantly and at random times. Our system does not interact with a thermal bath nor with a work source,
hence the energy exchanges is all quantum-heat, according to the definition introduced above. A direct consequence of previous studies~\cite{Campisi2010PRL,Campisi2011PRE} is that when the projective measurements occur at predetermined times, the Jarzynski equality of quantum-heat is obeyed. Here, we observe that the same is true when there is randomness in the waiting time distribution between consecutive measurements. In particular, we investigate both the cases when the randomness is distributed as a quenched disorder and as annealed disorder~\cite{Mezard1987}, for which we present the expression of the characteristic function. This can be understood based on the fact that the dynamics that dictate the evolution of the quantum system density matrix are unital~\cite{Albash13PRE88,Rastegin13JSTAT13,Kafri2012,Campisi17NJP19}.

Our general analysis is illustrated for a repeatedly measured two-level system, and we focus on the impact of randomness of waiting times on the average quantum-heat absorbed by the system. As compared with the case of measurements occurring at fixed times, the two-level system exchanges more quantum-heat in the presence of randomness, when the average time between consecutive measurements is sufficiently small compared to its inverse resonance frequency. Furthermore, more quantum-heat is absorbed by the two-level system when randomness is distributed as annealed noise as compared to quenched noise. Finally, quite surprisingly, we find that even an infinitesimal amount of randomness is sufficient to induce a non-null quantum-heat transfer when a large number of measurements on the system Hamiltonian is performed.

The manuscript is organized as follows. Section II introduces the protocol of stochastic measurement sequences performed on the quantum system, while in Section III the statistics of quantum-heat is derived, as well as the corresponding quantum Jarzynski equality. In Section IV we show the application of the previous analysis to the two-level
quantum system, providing analytical relations for the quantum-heat transfer. Finally, we discuss our results in Section V.

\section{\label{sec:model}II. Stochastic protocols}

We consider a quantum system $\mathcal{S}$ described by a finite dimensional Hilbert space $\mathcal{H}$. We assume that the system is initially at $t = 0^{-}$ in an arbitrary quantum state given a density matrix $\rho_0$. The system Hamiltonian $H$ is assumed to be time-independent and it reads:
\begin{equation}
H = \sum_{n}E_{n}|E_{n}\rangle\langle E_{n}|,
\label{eq:H}
\end{equation}
where $E_{n}$ and $|E_{n}\rangle$ are its eigenvalues and eigenstates, respectively. The eigenstates of $H$ are considered to be non-degenerate.

At time $t=0$ a first projective energy measurement occurs projecting the system in the state $\rho_{n} = \ket{E_{n}}\bra{E_{n}}$, with probability $p_{n} = \langle E_{n}|\rho_{0}|E_{n}\rangle$. Accordingly, the energy of the system at $t = 0^{+}$ is $E_{n}$. Afterwards, the system $\mathcal{S}$ is repeatedly subject to an arbitrary but fixed number $M$ of consecutive projective measurements of a generic observable $\mathcal O$
\begin{equation}\label{observable_o}
\mathcal{O} \equiv \sum_{k}o_{k}\Pi_{k}.
\end{equation}
Here, the $o_{k}$'s are the possible outcomes of the observable $\mathcal{O}$, while $\{\Pi_{k}\}$ is the set of the projectors belonging to the measured eigenvalues. The projectors are Hermitian and idempotent unidimensional operator satisfying the relations $\Pi_{k}\Pi_{l} = \delta_{kl}\Pi_{l}$ and $\sum_{k}\Pi_{k} = \mathbb{I}$. According to the postulate of quantum measurement~\cite{Sakurai1994}, the state of the quantum system after a projective measurement is given by one of the projectors $\Pi_{k}$.

We denote by $\tau_i$ the waiting time between the $(i-1)^\text{th}$ and the $i^\text{th}$ measurement of $\mathcal O$. Between those measurements the system undergoes the unitary dynamics generated by its Hamiltonian (\ref{eq:H}), that is $\mathcal{U}(\tau_i)= e^{-iH\tau_i}$, where the reduced Planck's constant $\hbar$ has been set to unity. The waiting times $\tau_i$ are random variables and so is the total time $\mathcal{T} = \sum_{i = 1}^{M}\tau_{i}$, when the last, i.e. the $M^\text{th}$, measurement of $\mathcal O$ occurs.

This is immediately followed by a second measurement of energy that projects the system on the state $\rho_{m} = \ket{E_{m}}\bra{E_{m}}$. The quantum-heat $Q_q$ absorbed by the system is accordingly given by:
\begin{equation}
Q_q = E_m -E_n.
\end{equation}
In the following we shall adopt the notation $\vec{\tau}=(\tau_{1},\ldots,\tau_{M})$ for the sequence of waiting time distributions, and $\vec{k}=(k_{1},\ldots,k_{M})$ for the sequence of observed outcomes from the measurement of $\mathcal O$ in a realisation of the measurement protocol. Given the sequences $\vec{k},\vec{\tau}$, the density matrix  $\rho_n$ is mapped at time $\mathcal T$ into
\begin{equation}\label{rho_tau_meno}
\widetilde{\rho}_{n,\vec{k},\vec{\tau}} = \frac{\mathcal{V}(\vec{k},\vec{\tau})\rho_{n}\mathcal{V}^{\dagger}(\vec{k},\vec{\tau})}
{\mathcal{P}(\vec{k},\vec{\tau})},
\end{equation}
where  $\mathcal{V}(\vec{k},\vec{\tau})$ is the super-operator
\begin{equation}\label{eq:ubar}
\mathcal{V}(\vec{k},\vec{\tau}) \equiv \Pi_{k_{M}}\mathcal{U}(\tau_M) \cdots \Pi_{k_{1}}\mathcal{U}(\tau_1)
\end{equation}
and $\mathcal{P}(\vec{k},\vec{\tau}) \equiv \Tr\left[ \mathcal{V}(\vec{k},\vec{\tau})\rho_{n}\mathcal{V}^{\dagger}(\vec{k},\vec{\tau}) \right]$.

\section{III. Quantum-heat statistics}

The quantum-heat $Q_{q}$ is a random variable due to the randomness inherent to the
measurements outcomes $\vec{k}$, the stochastic fluctuations in the sequence of waiting times $\vec{\tau}$, as well as to the initial statistical mixture $\rho_0$.
Its statistics reads
\begin{equation}\label{eq:pheat}
P(Q_{q}) = \sum_{n,m}\delta(Q_{q}-E_{m} + E_{n})p_{m|n}~p_{n},
\end{equation}
where $p_{m|n}$ is the transition probability to obtain the final energy $E_{m}$ conditioned to have measured $E_{n}$ in correspondence of the first energy measurement. Denoting with $p_{m|n}(\vec{k},\vec{\tau})$ the probability to make a transition from $n$ to $m$ conditioned on the waiting times and outcomes sequences $\vec{k},\vec{\tau}$, the overall transition probability $p_{m|n}$ reads
\begin{equation}\label{conditional_prob_an}
p_{m|n} = \int\sum_{\vec{k}}d^{M}\vec{\tau}p(\vec{\tau})p_{m|n}(\vec{k},\vec{\tau}),
\end{equation}
where $p(\vec{\tau})$ is the joint distribution of $\vec{\tau}$. Observe that the conditioned transition probability $p_{m|n}(\vec{k},\vec{\tau})$ is expressed in terms of the evolution super-operator $\mathcal{V}(\vec{k},\vec{\tau})$, so that
\begin{equation}
p_{m|n}(\vec{k},\vec{\tau}) = {\rm Tr}\left[\Pi_{m}\mathcal{V}(\vec{k},\vec{\tau})\Pi_{n}\mathcal{V}^{\dagger}(\vec{k},\vec{\tau})\Pi_{m}\right].
\end{equation}

The quantum-heat statistics is completely determined by the quantum-heat characteristic function
\begin{equation}\label{eq:G(u)}
G(u) \equiv \int P(Q_{q})e^{iuQ_{q}}dQ_{q},
\end{equation}
where $u\in\mathbb{C}$ is a complex number. Such characteristic function could be directly measured by means of Ramsey interferometry of single qubits~\cite{Dorner2013,MazzolaPRL2013,CampisiNJP2014}, or by means of methods from estimation theory~\cite{GherardiniEntropy}. Accordingly, plugging (\ref{conditional_prob_an}) into (\ref{eq:pheat}) the quantum-heat statistics becomes

\begin{small}
\begin{equation}\label{eq:pheat_2}
P(Q_{q}) = \int d^{M}\vec{\tau}  p(\vec{\tau})\sum_{n,\vec{k},m}{\rm Tr}\left[\Pi_{m}\mathcal{V}(\vec{k},\vec{\tau})\Pi_{n}\mathcal{V}^{\dagger}(\vec{k},\vec{\tau})\Pi_{m}\right]p_{n}.
\end{equation}
\end{small}

Moreover, as shown in the Appendix A1, substituting (\ref{eq:pheat_2}) in the definition (\ref{eq:G(u)}) and using ${\rm Tr}\left[\Pi_{m}\mathcal{V}\Pi_{n}\mathcal{V}^{\dagger}\Pi_{m}\right]=\bra{E_{m}}\mathcal{V}\ket{E_{n}}\bra{E_{n}}\mathcal{V}^{\dagger}\ket{E_{m}}$ we obtain
\begin{eqnarray}
G(u) &=& \int d^{M}\vec{\tau}  p(\vec{\tau})\sum_{n,\vec{k},m}
\bra{E_{m}}\mathcal{V}\ket{E_{n}}\bra{E_{n}}\rho_{0}\ket{E_{n}}\nonumber \\
&\cdot&\bra{E_{n}}e^{-iuH}\mathcal{V}^{\dagger}e^{iuH}\ket{E_{m}},
\end{eqnarray}
such that, being $e^{iuE_{m}}\ket{E_{m}} = e^{iuH}\ket{E_{m}}$ and $\bra{E_{n}}e^{-iuE_{n}} = \bra{E_{n}}e^{-iuH}$, the quantum-heat characteristic function reads
\begin{equation}\label{G_u_finale}
G(u) = \overline{\left\langle \Tr\left[e^{iuH}\mathcal{V}(\vec{k},\vec{\tau})e^{-iuH}\rho_{0}\mathcal{V}^{\dagger}(\vec{k},\vec{\tau})\right] \right\rangle}.
\end{equation}
In Eq. (\ref{G_u_finale}) the angular brackets mean quantum-mechanical expectation, i.e. $\langle \cdot \rangle = \Tr[ (\cdot) \rho]$, and the overline stands for the average over noise realisations: $\overline{(\cdot)} \equiv \int d^M\vec{\tau}p(\vec{\tau}) (\cdot)$.

In the special case when there is no randomness in the waiting times, i.e. if $p(\vec{\tau}) = \delta^M(\vec{\tau} - \vec{\tau}_0)$ where $\vec{\tau}_0 \equiv (\tau_0, \tau_0, \dots, \tau_0)$ and $\delta^M(\vec{x})$ denotes the $M$-dimensional Dirac delta, $G(u)$ reduces to
\begin{equation}\label{eq:G(u)_regular}
G(u) = \sum_{\vec{k}}
{\rm Tr}\left[e^{iuH}\mathcal{V}(\vec{k},\vec{\tau}_0)e^{-iuH}\rho_{0}\mathcal{V}^{\dagger}(\vec{k},\vec{\tau}_0)\right],
\end{equation}
in agreement with the expression in Ref.~\cite{Yi2013}.

The statistical moments of the quantum-heat are obtained from the derivatives of the quantum-heat generating function, according to the formula
\begin{equation}\label{work_moments}
\overline{\langle Q^{n}_{q}\rangle }=
\left.(-i)^{n}\partial^{n}_{u}G(u)\right|_{u=0},
\end{equation}
where $\partial^{n}_{u}$ denotes the $n-$th partial derivative with respect to $u$. Explicit expressions for $G(u)$ and $\overline{\langle Q^{n}_{q}\rangle}$ will be derived in the following section for the paradigmatic case of a two-level quantum system.

As a side remark we observe that, since the characterization of the measurement operators is encoded in the super-operator $\mathcal{V}(\vec{k},\vec{\tau})$, Eq.~(\ref{G_u_finale}) is valid also when a protocol of POVMs (excluding the first and the last measurements, performed on the energy basis) is applied to the quantum system. In such a case, the measurement projectors $\Pi_k$ are replaced by a set of Kraus operators $\{\mathcal{B}_{l}\}$, such that $\sum_{l}\mathcal{B}_{l}^{\dagger}\mathcal{B}_{l} = \mathbb{I}$.

\subsection{Fluctuation Relation}

It is known that, when a quantum system prepared in a thermal state with inverse temperature $\beta$ and subject to a time dependent forcing protocol and/or to a predetermined number of quantum projective measurements occurring at predetermined times $\vec{\tau}$, the following relation holds (Jarzynski equality):
\begin{equation}
\langle e^{-\beta (E'_m-E_n)}\rangle= e^{-\beta \Delta F},
\end{equation}
where $E'_m$ are the final eigenvalues of the time-dependent system Hamiltonian $H(t)$, $\Delta F \equiv - \beta^{-1}\ln \Tr[e^{-\beta H(\mathcal T)}]/ \Tr[e^{-\beta H(0)}]$ denotes the free-energy difference, and the initial state of the system has the Gibbs form $\rho_0=e^{-\beta H(0)} / \Tr[e^{-\beta H(0)}]$ \cite{Campisi2011PRE}. $Z \equiv \Tr[e^{-\beta H(0)}]$ is also called partition function. If the time-dependent forcing is turned off, as in the present investigation, then
with \textit{fixed} waiting times $\vec{\tau}$ one has:
\begin{equation}\label{eq:av-exp(-Qq)}
\langle e^{-\beta Q_q}\rangle = 1,
\end{equation}
because, without driving, all the energy change in the quantum system can be ascribed to quantum-heat and, being the Hamiltonian time-independent, in that case $\Delta F = 0$. For the sake of clarity, we recall that the notation $\langle e^{-\beta Q_q}\rangle$ denotes a purely quantum-mechanical expectation with fixed waiting time sequence $\vec{\tau}$.

However, as main result, we can easily prove that this continues to hold also if the times between consecutive measurements are random. Indeed, using (\ref{G_u_finale}), we obtain
\begin{align}
&\overline{\langle e^{-\beta Q_q} \rangle} = G(i\beta)\nonumber \\
& = \int d^{M}\vec{\tau}  p(\vec{\tau})\sum_{\vec{k}}
{\rm Tr}\left[e^{-\beta H}\mathcal{V}(\vec{k},\vec{\tau})e^{\beta H}\frac{e^{-\beta H}}{Z}\mathcal{V}^{\dagger}(\vec{k},\vec{\tau})\right] \nonumber \\
&={\rm Tr}\left[\frac{e^{-\beta H}}{Z} \int d^{M}\vec{\tau}  p(\vec{\tau}) \sum_{\vec{k}}\mathcal{V}(\vec{k},\vec{\tau})\mathcal{V}^{\dagger}(\vec{k},\vec{\tau})\right]\nonumber \\
&= \frac{{\rm Tr}\left[e^{-\beta H}\right]}{Z} =1,
\label{G_u_finale_appendice}
\end{align}
where we have used the property
\begin{align}
\int d^{M}\vec{\tau}p(\vec{\tau}) \sum_{\vec{k}}\mathcal{V}(\vec{k},\vec{\tau})\mathcal{V}^{\dagger}(\vec{k},\vec{\tau})=\mathbb{I},
\end{align}
which follows from the normalisation $\int d^{M}\vec{\tau}  p(\vec{\tau})=1$, idempotence of projectors $\Pi_k\Pi_k=\Pi_k$, cyclicity of the trace operation, and the unitarity of the quantum evolutions between consecutive measurements. Its mathematical significance is that the quantum channel that describes the unconditioned evolution from $t=0$ to $t=\mathcal T$
\begin{align}
\rho \mapsto \int d^{M}\vec{\tau}  p(\vec{\tau}) \sum_{\vec{k}}\mathcal{V}(\vec{k},\vec{\tau})\, \rho \, \mathcal{V}^{\dagger}(\vec{k},\vec{\tau})
\end{align}
is \emph{unital}, i.e. it has the identity $\mathbb{I}$ as a fixed point. It is this property that ensures the validity of the fluctuation relation (\ref{G_u_finale}) \cite{Albash13PRE88,Rastegin13JSTAT13,Kafri2012,Campisi17NJP19}.

The fluctuation relation (\ref{G_u_finale}) can also be understood by noticing that, from Eq. (\ref{eq:av-exp(-Qq)}), $\langle e^{-\beta Q_q}\rangle = 1$, in which the average is restricted to the sole realisations where the sequence $\vec{\tau}$ occurs. The double average remains therefore equal to one:
$\overline{\langle e^{-\beta Q_q} \rangle} = \int d^M{\vec \tau} p(\vec{\tau})\langle e^{-\beta Q_q}\rangle=1$. Accordingly, we have shown, from one side, that the fluctuation relation is \textit{robust} against the presence of randomness in the waiting times $\vec{\tau}$, and, on the other side, that such stochasticity shall not be a-posterior revealed by a measure of $\overline{\langle e^{-\beta Q_q} \rangle}$ with $\rho_0$ Gibbs thermal state, whatever are the values assumed by $\vec{\tau}$ and $p(\vec{\tau})$.

Finally, from an experimental point of view, $\overline{\langle e^{-\beta Q_{q}}\rangle}$ could be obtained by repeating for a sufficiently large number $N$ of times the foregoing protocol of projective measurements, so that
\begin{equation}
\overline{\langle e^{-\beta Q_{q}}\rangle} = \lim_{N \to \infty} \frac{1}{N}\sum_{j=1}^{N}e^{-\beta Q_{q}^{(j)}},
\label{traj}
\end{equation}
where $Q_{q}^{(j)}$ is the value of quantum-heat, which is measured after the $j-$th repetition of the experiment. It is worthwhile to observe that, according the previous discussion, if not all the realizations of the distribution of the waiting times are present in the experimental measurements, but $N$ is large enough, the Jarzynski equality given by the average (\ref{traj}) on the system trajectories is anyway satisfied.

\section{IV. Quantum-heat transfer by a two-level-system}

In this Section, we analyze in detail $\overline{\langle e^{-\beta Q_{q}}\rangle}$ and the mean quantum-heat $\overline{\langle Q_{q}\rangle}$, when a \textit{stochastic} sequence of projective quantum measurements is performed on a two-level system. Let $E_{+}$ and $E_{-}$ denote its two energy eigenvalues. Being the first measurement of the protocol an energy measurement, we assume without loss of generality that the initial density matrix is diagonal in the energy eigenbasis:
\begin{equation}\label{eq:rho_0_2_level_system}
\rho_{0} = c_{1}\ket{E_{+}}\bra{E_{+}} + c_{2}\ket{E_{-}}\bra{E_{-}},
\end{equation}
with $c_1,c_2 \in[0,1]$ and $c_2=1 - c_1$, and we denote the eigenstates of the measured observable $\mathcal O$ as $\{|\alpha_{j}\rangle\}$, $j = 1,2$, so that is $\Pi_j = |\alpha_{j}\rangle \langle \alpha_j|$. The projectors could be generally expressed as a linear combination of the energy eigenstates, i.e.
\begin{equation}
\begin{split}
&\ket{\alpha_1}=a\ket{E_{+}}-b\ket{E_{-}}\\
&\ket{\alpha_2}=b\ket{E_{+}}+a\ket{E_{-}}
\label{eq:abasis}
\end{split}
\end{equation}
where $a,b\in\mathbb{C}$, $|a|^2+|b|^2=1$ and $a^{\ast}b = ab^{\ast}$.

\subsection{Fixed waiting times sequence}

We begin by considering the standard case where the waiting time $\overline{\tau}$ between two consecutive measurements is constant. In this case
\begin{equation}
p(\vec{\tau}) = \prod_{i=1}^M \delta(\tau_i-\bar\tau).
\end{equation}
By computing the characteristic function (\ref{eq:G(u)_regular}) in $u = i\beta$, we obtain
\begin{eqnarray}\label{G_i}
G(i\beta)&=&\begin{pmatrix}
|a|^{2}e^{-\beta E}+|b|^{2}e^{\beta E}\\
|a|^{2}e^{\beta E} + |b|^{2}e^{-\beta E}
\end{pmatrix}^{T}\cdot\begin{pmatrix}
1-\overline{\nu} & \overline{\nu} \\ \overline{\nu} & 1-\overline{\nu}
\end{pmatrix}^{M-1}\nonumber \\
&\cdot&\begin{pmatrix}
|a|^{2}c_{1}e^{\beta E}+|b|^{2}c_{2}e^{-\beta E}\\
|a|^{2}c_{2}e^{-\beta E} + |b|^{2}c_{1}e^{\beta E}
\end{pmatrix},
\end{eqnarray}
where the transition probability $\overline{\nu} = \nu(\overline{\tau})$ is expressed in terms of the function
\begin{eqnarray}
\nu(t) &\equiv& |\bra{\alpha_{2}}\mathcal{U}(t)\ket{\alpha_{1}}|^2=|\bra{\alpha_{1}}\mathcal{U}(t)\ket{\alpha_{2}}|^2\nonumber \\
&=& 2|a|^{2}|b|^{2}\sin^{2}(2 t E).
\end{eqnarray}
The explicit calculation is reported in the Appendix A2.

\begin{figure}[t!]
\includegraphics[scale = 0.482]{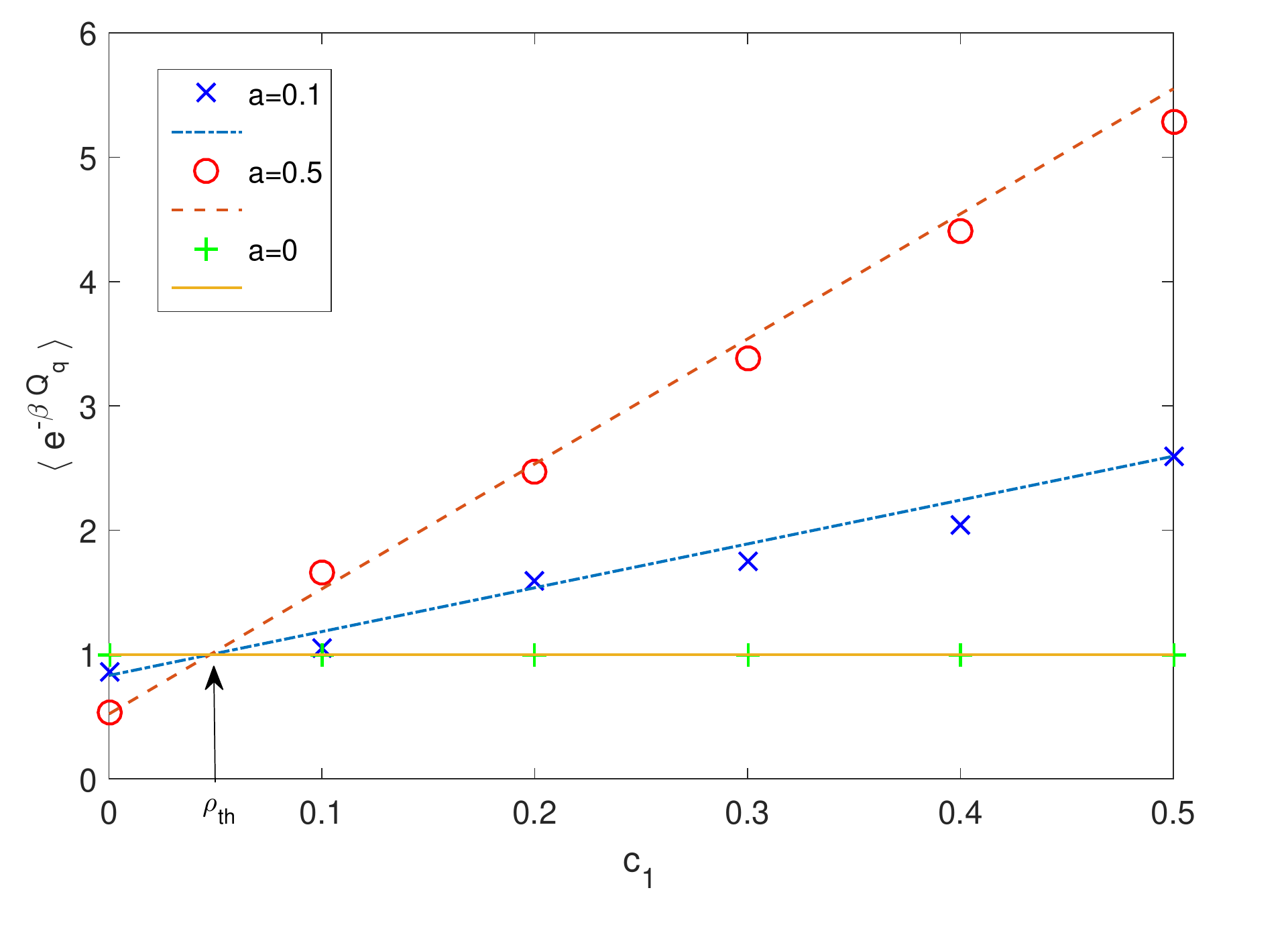}
\caption{Analytical result for $G(i\beta)=\langle e^{-\beta Q_q}\rangle$ for a fixed waiting times sequence as a function of $c_{1}$ for three real values of $a$ (respectively, $a = 0, 0.1, 0.5$ - solid yellow, dotted blue and dashed red lines). The analytical predictions are compared with the numerical simulations (green crosses, blue x-marks, and red circles). The simulations have been performed by applying protocols of $M = 5$ projective measurements, averaged over $1000$ realizations in order to numerically derive the mean of the exponential of work, with $E_{\pm} = \pm 1$. The point in which all the analytical lines are crossing corresponds to take the initial density matrix of the system equal to the thermal state $\rho_{{\rm th}}=e^{-\beta H}/Z$ with $\beta = 1$.}
\label{fig:QJI}
\end{figure}
In Fig. \ref{fig:QJI} we plot the quantity $G(i\beta)= \langle e^{-i\beta Q_q} \rangle$ as a function of $c_1$ for various values of $a$, which have been chosen to be real. We first observe that $G(i\beta)$ is a linear function of $c_1$. This is confirmed by the numerical simulations of $\langle e^{-i\beta Q_q} \rangle $ from the underlying protocol, which is in agreement with the analytical formula (\ref{G_i}), except for some finite size errors. We further observe that, for an arbitrary value of $a$, $G(i\beta)$ is identically equal to $1$ in correspondence of the value of $c_{1}$ for which $\rho_0=e^{-\beta H}/Z$, in agreement with Eq.~(\ref{eq:av-exp(-Qq)}). In Fig.~\ref{fig:QJI} such condition is realized in the point where all the analytical lines are crossing.

\subsection{Stochastic waiting times sequence}

\subsubsection{Quenched disorder}

We consider here the case in which the time between consecutive measurements within a given sequence is fixed and only varies between distinct sequences. In other words, only the first waiting time of each sequence is chosen randomly from $p(\tau)$ and, then, it is identically repeated $M$ times within the sequence. Accordingly, the joint distribution $p(\vec{\tau})$ reads
\begin{equation}
p(\vec{\tau}) = p(\tau_1) \prod_{i=2}^M \delta(\tau_i-\tau_1).
\end{equation}
For the sake of simplicity, we assume that $p(\tau)$ is a bimodal probability density function, with values $\tau^{(1)}$, $\tau^{(2)}$ and probabilities $p_{1}$ and $p_{2} = 1 - p_{1}$. Accordingly, from Eq. (\ref{G_u_finale}) we have

\begin{footnotesize}
\begin{eqnarray}\label{eq:G_quenched_2_LS}
&G(i\beta)=&\nonumber \\
&\begin{pmatrix}
|a|^{2}e^{-\beta E}+|b|^{2}e^{\beta E}\\
|a|^{2}e^{\beta E} + |b|^{2}e^{-\beta E}
\end{pmatrix}^{T}\cdot\left[\displaystyle{\sum_{j = 1}^{d_{\tau}}}
\begin{pmatrix}
1-\nu(\tau^{(j)}) & \nu(\tau^{(j)}) \\
\nu(\tau^{(j)}) & 1-\nu(\tau^{(j)})
\end{pmatrix}^{M-1}p_{j}\right]&\nonumber \\
&\cdot\begin{pmatrix}
|a|^{2}c_{1}e^{\beta E}+|b|^{2}c_{2}e^{-\beta E}\\
|a|^{2}c_{2}e^{-\beta E} + |b|^{2}c_{1}e^{\beta E}
\end{pmatrix}&
\end{eqnarray}
\end{footnotesize}

where $d_{\tau} = 2$ is the number of values that can be assumed by the random variable $\tau$.

\subsubsection{Annealed disorder}

For an annealed disorder we assume that the waiting times $(\tau_{1},\ldots,\tau_{M})= \vec{\tau}$ are random variables sampled from the same probability distribution $p(\tau)$. Accordingly, the joint distribution of the waiting times is
\begin{equation}
p(\vec{\tau}) = \prod_{j=1}^{M}p(\tau_{j}).
\end{equation}
Assuming $p(\tau)$ to be bimodal as above, the characteristic function at $u=i\beta$ reads (see Appendix A2):
\begin{footnotesize}
\begin{eqnarray}\label{eq:G_annealed_2LS}
&G(i\beta)=&\nonumber \\
&\begin{pmatrix}
|a|^2e^{-\beta E}+|b|^2e^{\beta E}\\
|a|^2e^{\beta E}+|b|^2e^{-\beta E}
\end{pmatrix}^{T}\cdot\left[\displaystyle{\sum_{j = 1}^{d_{\tau}}}\begin{pmatrix}
1-\nu(\tau^{(j)}) & \nu(\tau^{(j)}) \\ \nu(\tau^{(j)}) & 1-\nu(\tau^{(j)})
\end{pmatrix}p_{j}\right]^{M-1}&\nonumber \\
&\cdot\begin{pmatrix}
|a|^2c_{1}e^{\beta E}+|b|^2c_{2}e^{-\beta E}\\
|a|^2c_{2}e^{-\beta E}+|b|^2c_{1}e^{\beta E}
\end{pmatrix}&
\end{eqnarray}
\end{footnotesize}

\begin{figure}[t]
\includegraphics[scale = 0.56]{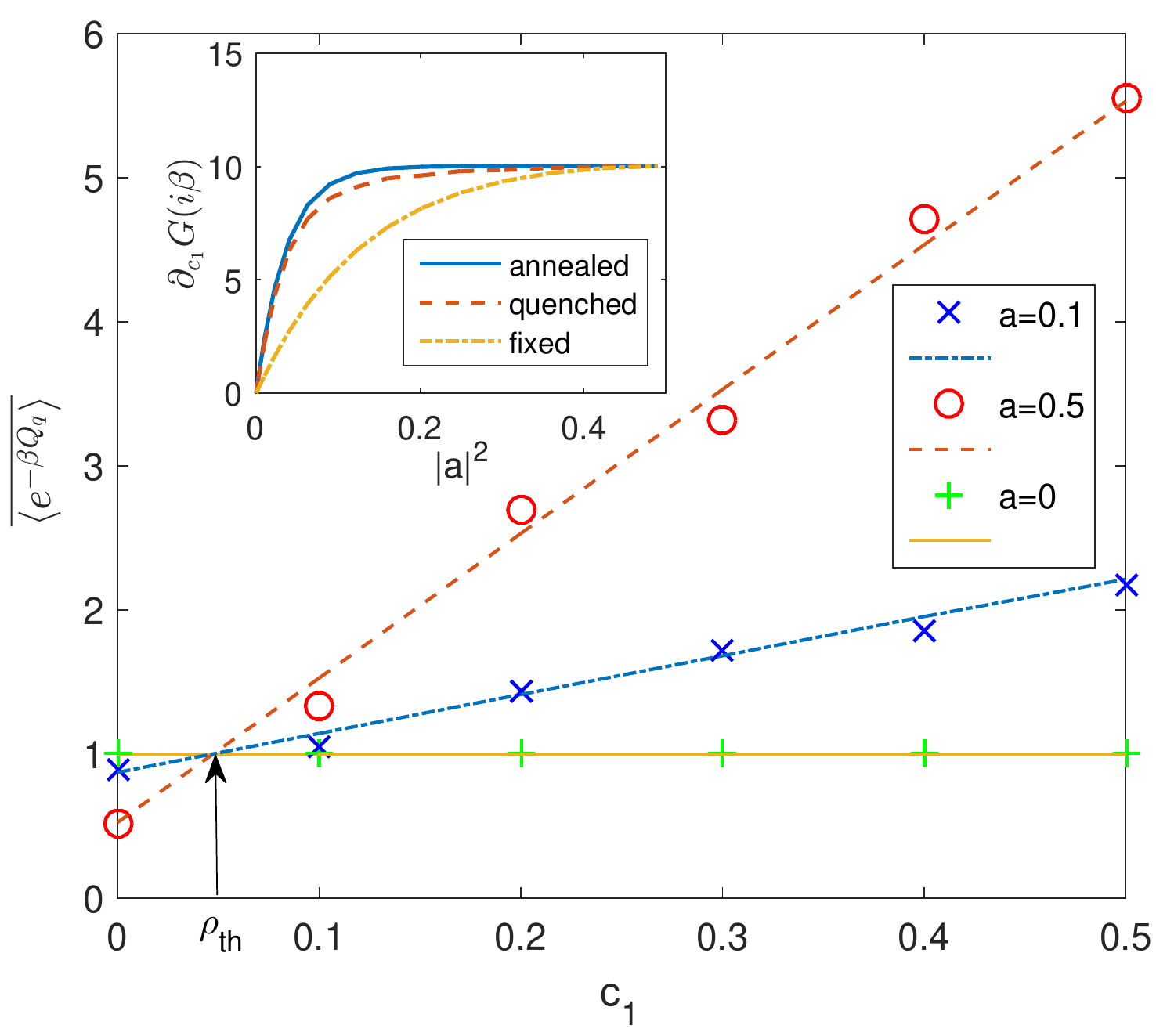}
\caption{$G(i\beta) = \overline{\langle e^{-\beta Q_q}\rangle}$ (solid yellow, dotted blue and dashed red lines) as a function of $c_{1}$ for three real values of $a$ ($a = 0, 0.1, 0.5$, respectively). The stochasticity in the time intervals between measurements is considered to be distributed as an annealed disorder. The analytical prediction is compared to the numerical simulations (green crosses, blue x-marks and red circles) for the three values of $a$. Also in this case, the point in which all the lines are crossing corresponds to the thermal state. Inset: Partial derivative $\partial_{c_1}G(i\beta)$ as a function of the measurement operator parameter $|a|^2$, with resolution of $|a| = 0.05$, for three different types of noise (fixed, quenched and annealed). The curves have been performed by applying protocols of $M = 5$ projective measurements, averaged over $1000$ realizations, with $E_{\pm} = \pm 1$ and $\beta = 1$. For $p(\tau)$ we have chosen a bimodal probability density function, with values $\tau^{(1)}=0.01$, $\tau^{(2)}=3$ and $p_{1}=0.3$.}
\label{fig:QJIrand}
\end{figure}
In Fig.~\ref{fig:QJIrand} we plot it as a function of $c_{1}$. The presence of the disorder does not affect the linear dependence of $G(i\beta)$ on  $c_1$, and it still equals to $1$ in correspondence of the initial state being thermal with temperature $1/\beta$. What the stochasticity effectively changes is the sensitivity of $G(i\beta)$ with respect to the initial state, parameterized by $c_1$. In this regard, in the inset of Fig.~\ref{fig:QJIrand} we show how the partial derivative $\partial_{c_1}G(i\beta)$ depends on the parameter $|a|^2$ that defines the measurement operator. We show results for fixed and stochastic (quenched and annealed) waiting time sequences with $M=5$ measurements. The values of $\partial_{c_1}G(i\beta)$ are identically equal when $|a|^2 = 0$ and $1/2$, and in the range $0 \leq |a|^2 \leq 1/2$ they are symmetric with respect to the ones in the range $1/2 \leq |a|^2 \leq 1$.

\subsection{Mean quantum-heat}

In the Appendix A2 we show the analytical expression of the $n-$th partial derivative of $G(u)$ for the two-level system. By substituting $u = 0$ in $\partial_{u}G(u)$, we find the mean value $\overline{\langle Q_{q}\rangle}$, which is a linear function in the parameter $c_1$ both in the ordered and the stochastic case. In particular,
\begin{equation}
\overline{\langle Q_{q}\rangle} = -\phi\left[2c_1 - 1\right],
\label{eq:heat_normal}
\end{equation}
where $\phi \equiv E\left[1 - \overline{\lambda(\tau)}\right]$, and $\lambda(\tau)$ is given by the following relation:
\begin{equation}
\lambda(\tau) = (1 - 2|a|^{2})^{2}(1 - 2\nu(\tau))^{M-1}\leq 1.
\end{equation}
As it can be observed, $\phi$ depends on the average of the parameter $\lambda(\tau)$ w.r.t. the values that the waiting time $\vec{\tau}$ can assume in a given sequence of the protocol according to $p(\vec{\tau})$. Therefore, the (linear) dependence of the mean quantum-heat as a function of $c_1$ varies for different values of $\tau$ and corresponding probabilities. Being $\phi\geq 0$, the maximum value of $\overline{\langle Q_{q}\rangle}$, i.e. $\overline{\langle Q_{q}\rangle}_{\rm max}$, occurs at $\overline{\langle Q_{q}\rangle} = \phi$ when $c_1 = 0$, while $\overline{\langle Q_{q}\rangle} = 0$ when $c_1 = 1/2$ for any value of $M$, $a$ and $p(\vec{\tau})$. Moreover, $\overline{\langle Q_{q}\rangle} = 0$ also when $a=0$ or $a=1$. This can be understood by noticing that the condition $a=0,1$ implies that the measured observable $\mathcal O$ coincides with the system Hamiltonian. In this case the system after the initial projection onto the state $|E_{\pm}\rangle$ only acquires a phase during the free evolution while the subsequent measurements have no effect on the state. Thus, the quantum-heat is always vanishing ($Q_{q} = 0$), and so it will be its average.

For a sequence of measurements at fixed times one has $\overline{\lambda(\tau)} = \lambda(\overline{\tau})$, while for quenched and annealed disorder the average of $\lambda(\tau)$ is respectively equal to
\begin{eqnarray}
\overline{\lambda(\tau)}^{(\rm qu)} &=& \sum_{j = 1}^{d_{\tau}}\lambda(\tau^{(j)})p_j\nonumber \\
&=& (1 - 2|a|^2)^{2}\sum_{j = 1}^{d_{\tau}}[1 - 2\nu(\tau^{(j)})]^{M-1}p_j
\end{eqnarray}
and
\begin{equation}
\overline{\lambda(\tau)}^{(\rm an)} = (1 - 2|a|^2)^{2}\left[\displaystyle{\sum_{j=1}^{d_{\tau}}}[1-2\nu(\tau^{(j)})]p_{j}\right]^{M-1}.
\end{equation}
We will denote the mean quantum-heat in such cases respectively as $\overline{\langle Q_{q}\rangle}^{(\rm qu)}$ and $\overline{\langle Q_{q}\rangle}^{(\rm an)}$.

By changing the initial density matrix $\rho_{0}$ (i.e. $c_1$), the parameter $a$ (related to the measurement bases) or the number $M$ of measurements, the mean value of the quantum-heat assumes a value within the range $[-\phi,\phi]$. In particular, when the initial state is thermal, then $\overline{\langle Q_{q}\rangle} = \beta E(1 - \overline{\lambda(\tau)})\tanh(\beta E)$, as also shown in Ref. \cite{Yi2013} for a sequence of measurements at fixed times. Moreover, we observe that $\overline{\langle Q_{q}\rangle} \geq 0$ if $0\leq c_1\leq 1/2$, while it is $\overline{\langle Q_{q}\rangle} \leq 0$ for $1/2\leq c_1\leq 1$. These two conditions correspond to two distinct regimes for the two-level system: quantum-heat absorption, and quantum-heat emission. Then, since $\overline{\langle Q_{q}\rangle}$ is a linear function passing through $c_1 = 1/2$, the quantum-heat transfer (heat absorption/emission) can be studied just by comparing the absolute value of the maximum quantum-heat, i.e. $\left|\overline{\langle Q_{q}\rangle}_{\rm max}\right| = \phi$, for sequences of measurements at fixed and stochastic times. This implies to determine the relations between $\lambda(\overline{\tau})$, $\overline{\lambda(\tau)}^{(\rm qu)}$ and $\overline{\lambda(\tau)}^{(\rm an)}$. We find that
\begin{equation}\label{inequalities_quenched}
\left|\overline{\langle Q_{q}\rangle}^{({\rm qu})}\right| \geq \left|\langle Q_{q}\rangle\right| \Longleftrightarrow (1-2\nu)^{M-1} \geq \overline{[1-2\nu(\tau)]^{M-1}}^{({\rm qu})}.
\end{equation}
and
\begin{equation}
\left|\overline{\langle Q_{q}\rangle}^{({\rm an})}\right| \geq \left|\overline{\langle Q_{q}\rangle}^{({\rm qu})}\right|,
\end{equation}
being $\overline{\lambda(\tau)}^{(\rm an)} \leq \overline{\lambda(\tau)}^{(\rm qu)}$. Eq.~(\ref{inequalities_quenched}) sets the condition allowing for the transfer on average of a greater amount of quantum-heat under the case of quenched noise as compared to the case of no noise. To better understand its physical meaning, let us consider $\tau^{(j)}\Delta E \ll 1$, $j = 1,2$. One has
\begin{equation}\label{inequalities_quenched_2}
\left|\overline{\langle Q_{q}\rangle}^{({\rm qu})}\right| \geq \left|\langle Q_{q}\rangle\right| \Longleftrightarrow \overline{\tau}^{2} \geq \overline{\tau^2},
\end{equation}
where $\overline{\tau^2}$ is the second statistical moment of $p(\tau)$. If the condition (\ref{inequalities_quenched_2}) is not verified, then the application of a sequence of measurements at fixed times leads to a greater amount of transferred quantum-heat. Instead, for a given choice of $p(\tau)$ and total number of measurements $M$, more quantum-heat is absorbed/emitted by the two-level system in the case of annealed noise as compared to the quenched noise case. This agrees with the intuition that the system should heat-up more in case it is subjected to higher noise, and the annealed disorder is ``noiser'' than the quenched one. This evidences a phenomenon of \textit{noise-induced quantum-heat transfer}, which we will be investigating further elsewhere.

As final remark, it is worth mentioning that in recent studies on stochastic quantum Zeno dynamics \cite{Gherardini2017QSc,Q_Zeno_noise_detection}, it has been shown that the survival probabilities that the system remains frozen in its initial state after performing ordered and stochastic sequences of measurements behave in the opposite way. In other words, the better the Zeno confinement is, the less quantum-heat is transferred by the system.

\subsection{The $M\rightarrow \infty$ limit}

For $M\rightarrow\infty$ the characteristic function tends to $G_\infty(u) = (1+e^{2iuE})/2 - c_{1}\sinh(2iuE)$ for each value of $a\neq 0$ and it is exactly equal to $1$ for $|a|^2=0,1$. Therefore the $M\rightarrow \infty$ asymptotic characteristic function $G_\infty(u)$ presents a discontinuity at $|a|^2=0,1$. Such a discontinuity is present also in the mean quantum-heat $\overline{\langle Q_{q}\rangle}$. When $|a|^2\rightarrow 0,1$ and $M$ is finite, $\overline{\langle Q_{q}\rangle}\rightarrow 0$ for any value of $c_1$, while for $M\rightarrow\infty$ and $|a|^2 \neq 0,1$ we get $\overline{\langle Q_{q}\rangle}\rightarrow E(1 - 2c_1) = \overline{\langle Q_{q}\rangle}_{\infty}$. In this way, the $M\rightarrow \infty$ asymptotic mean quantum-heat $\overline{\langle Q_{q}\rangle}_\infty$ can be easily expressed in terms of the $M \rightarrow \infty$ asymptotic characteristic function $G_\infty(u)$:
\begin{equation}
G_\infty(u) = \frac{\sinh(2iuE)}{E}\overline{\langle Q_{q}\rangle}_\infty + [\cosh(2iuE) + 1].
\end{equation}

The existence of this discontinuity is a mathematical feature that is physically relevant when one performs many measurements ($M\rightarrow\infty$) of the Hamiltonian ($|a|^2\rightarrow 0,1$). Perfect measurements of the Hamiltonian ($|a|^2\rightarrow 0,1$) are accompanied by vanishing quantum-heat $\overline{\langle Q_{q}\rangle}$; however even an infinitesimal amount of noise in the measurement process results, in the limit of a very large number of measurements, in a finite amount of quantum-heat $\overline{\langle Q_{q}\rangle}_\infty=E(1 - 2c_1)$. Note that the latter is positive (negative) if the initial state is at positive (negative) temperature, i.e. $c_1 >c_2$ ($c_1 <c_2$).

\subsection{Comparing the protocols at fixed $\mathcal{T}$}

In the following we assume to fix the total time of the protocol and the mean waiting time $\langle\tau\rangle$. In particular, we impose that $\langle\tau\rangle = \overline{\tau}$. Fixing the total time $\mathcal{T}$ and the mean waiting time to $\overline{\tau}$ allows us to compare between them the results about quantum-heat absorption/emission (according to the value of $c_{1}$), which are obtained by applying a sequence of measurements at stochastic or fixed waiting times within the same time interval, as it could be reasonably done in an experimental setup. In other words, we aim to observe if and when the quantum-heat transfer is enhanced by the presence of the stochasticity in the protocol.

Note that, if we fix the total time of the protocol, the number $M$ of measurements becomes a random variable, fluctuating around the corresponding mean value $\langle M\rangle$. In the limiting case of $\langle M\rangle\rightarrow\infty$, by comparing Eq. (\ref{eq:heat_normal}) in the ordered and stochastic cases, we find in the range $0 \leq c_1 \leq 1/2$ that
\begin{equation}
\langle Q_{q}\rangle_{{\rm max}} = \overline{\langle Q_{q}\rangle}_{{\rm max}}^{({\rm qu})} = \overline{\langle Q_{q}\rangle}_{{\rm max}}^{({\rm an})} = 4E|a|^{2}(1-|a|^{2}),
\end{equation}
whatever is the value of $\langle\tau\rangle$. In particular, when $|a|^{2} = 1/2$, the mean quantum-heat reaches the maximum value $E$.
\begin{figure}[h!]
\includegraphics[scale = 0.52]{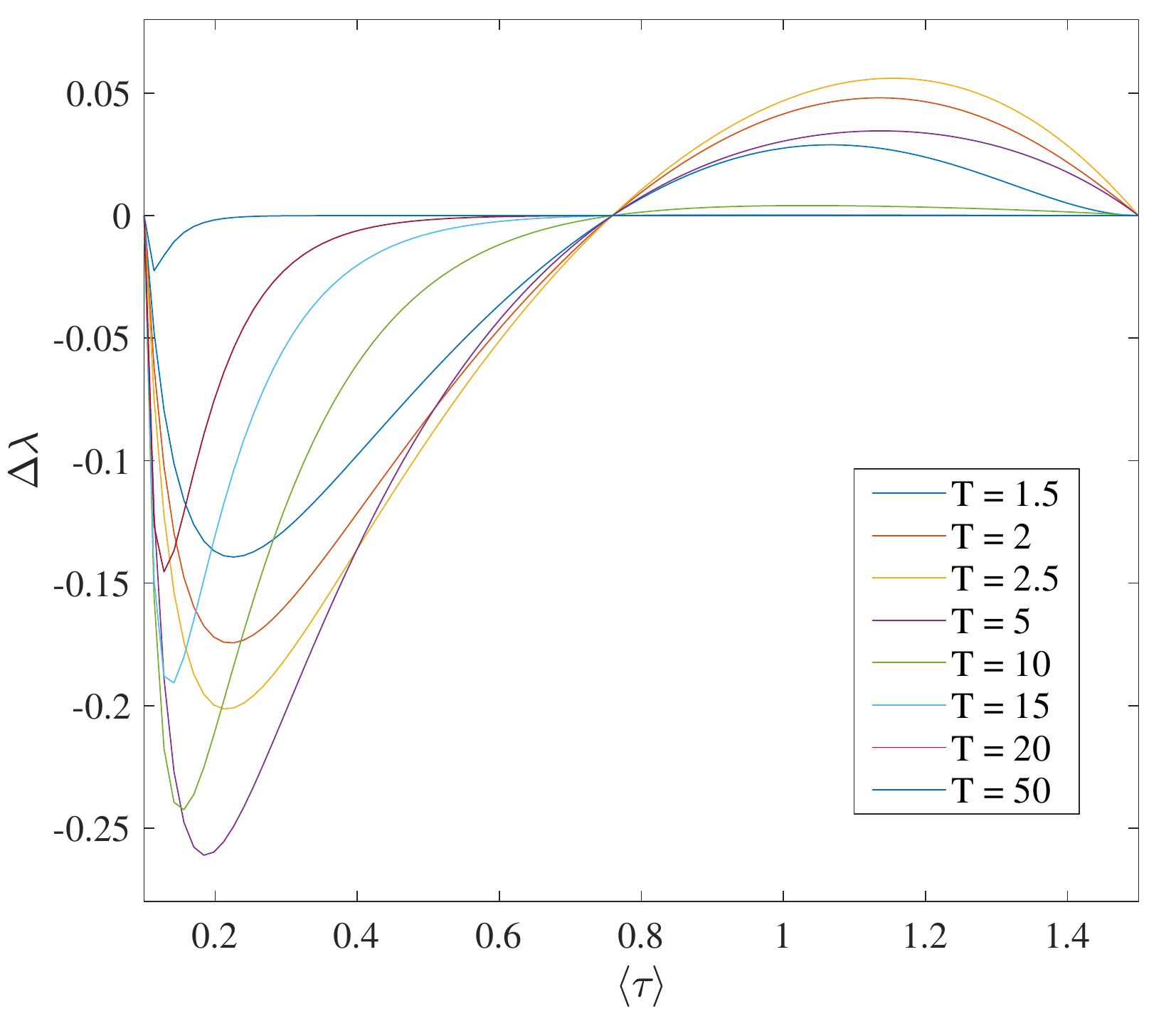}
\caption{$\Delta\lambda$ as a function of $\langle\tau\rangle = \overline{\tau}$ for different values of
$\mathcal{T}=1.5,2,2.5,5,10,15,20,50$. By analyzing $\Delta\lambda$, we are able to compare the maximum value of the mean quantum-heat which is transferred by a two-level quantum system subject to different protocols of projective measurements. In the numerics, we have chosen $\Delta E=1$, $\tau^{(1)} = 0.1$, $\tau^{(2)} = 1.5$ and $|a|^2 = 0.2$.}
\label{fig:heat_absorption_2}
\end{figure}
By defining $\Delta\lambda \equiv \lambda - \langle\lambda\rangle^{({\rm an})}$, in Fig. \ref{fig:heat_absorption_2} we plot $\Delta\lambda$ as a function of $\langle\tau\rangle$ by choosing $\tau^{(1)} = 0.1$ and $\tau^{(2)} = 1.5$ with $\Delta E = 1$. $\Delta\lambda$ compares $\langle Q_{q}\rangle_{{\rm max}}$ with $\overline{\langle Q_{q}\rangle}_{{\rm max}}^{({\rm an})}$, and, particularly, $\Delta\lambda < 0$ implies an enhancement of the mean quantum-heat transfer under the effects of a stochastic protocol. In Fig. \ref{fig:heat_absorption_2}, when the range of values of $\langle\tau\rangle$ is smaller than the energy splitting $\Delta E \equiv E_{+} - E_{-}$, $\Delta\lambda < 0$ for a wide set of different final times $\mathcal{T}$. This property, valid when $\langle\tau\rangle\Delta E \ll 1$,
can be considered as a further peculiarity of protocols given by sequences of measurements in the recently introduced stochastic Zeno regime \cite{Gherardini2016NJP}. Therefore, although there exist cases in which $\Delta\lambda$ as a function of $\langle\tau\rangle$ is always smaller than zero (e.g. by choosing $\tau^{(1)} = 0.1$, $\tau^{(2)} = 0.5$ and $\Delta E = 1$), in general the application of a stochastic protocol of measurements does not involve a stronger absorption/emission of quantum-heat with respect to those at fixed waiting times, and, thus, they could be employed to carry out specific control tasks on the quantum system. As additional remark, let us note that, being $\langle\lambda\rangle^{({\rm an})}\leq \langle\lambda\rangle^{({\rm qu})}$, if $\Delta\lambda\leq 0$ then also $\lambda \leq \langle\lambda\rangle^{({\rm qu})}$; hence, these considerations are also valid in the quenched disorder case.

\begin{figure}
\includegraphics[scale = 0.6]{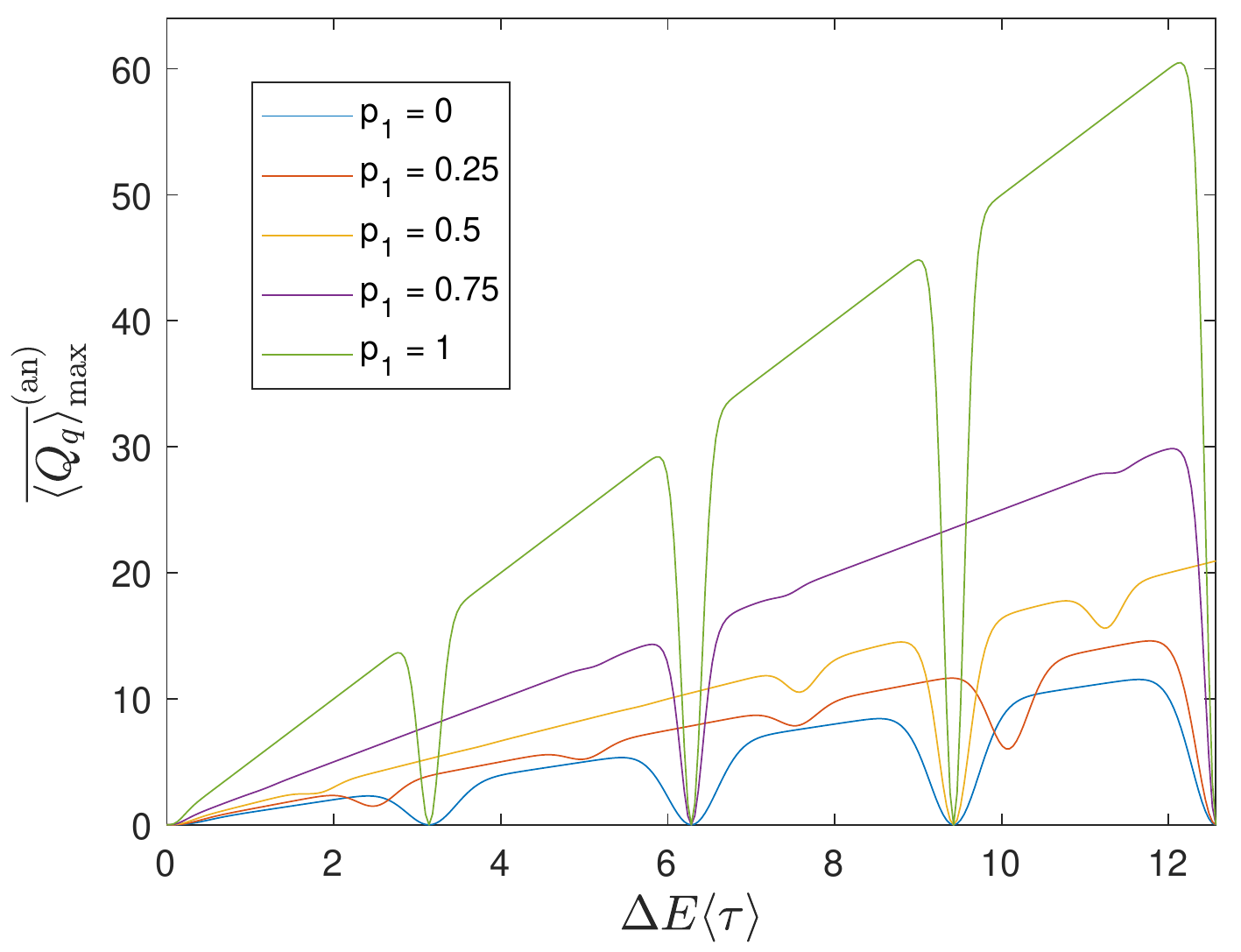}
\caption{Maximum mean quantum-heat $\overline{\langle Q_{q}\rangle}_{\text{max}}^{({\rm an})}$ as a function of $\Delta E\langle\tau\rangle \in [0,4\pi]$, when a stochastic sequence of measurements (annealed disorder) is applied to a two-level system. The curves have been obtained by varying $p_1$ among the set of values $(0,0.25,0.5,0.75,1)$, and by choosing $\mathcal{T} = 5$, $\tau^{(1)} = 0.1$, $\tau^{(2)} = 0.5$ and $|a|^2 = 0.2$.}
\label{fig:mean_q_heat}
\end{figure}

In Fig. \ref{fig:mean_q_heat}, we show the behavior of the maximum value of the mean quantum-heat $\overline{\langle Q_{q}\rangle}_{\text{max}}^{({\rm an})}$ as a function of $\Delta E\langle\tau\rangle$ in the range $0 \leq c_1 \leq 1/2$. Notice that if $p_1 = 0$ or $1$, then the results will have to be referred to a sequence of measurements with fixed waiting times with $\overline{\tau}$, respectively, equal to $\tau^{(1)}$ and $\tau^{(2)}$. In this latter case, $\overline{\langle Q_{q}\rangle}_{\text{max}}^{({\rm an})}$ is zero when $\Delta E\langle\tau\rangle = n\pi$, so as to exhibit a resonance in such points. The reason is that the dynamics between consecutive measurements, being defined by an unitary evolution operator, is periodic with period $2\pi/\Delta E $. However, by introducing stochastic contributions in the waiting times between measurements ($p_1\neq 0,1$), such periodicity gets lost, as well as the resonant behavior of $\overline{\langle Q_{q}\rangle}_{\text{max}}^{({\rm an})}$. The only exception, as shown in the figure, is given when the ratio between $p_1$ and $p_2$ is commensurable with the ratio between $\tau^{(1)}$ and $\tau^{(2)}$. Hence, to conclude, in correspondence of the points where $\Delta E\langle\tau\rangle = n\pi$ one has $\Delta\lambda < 0$, so that $\overline{\langle Q_{q}\rangle}_{{\rm max}}^{({\rm an})} > \langle Q_{q}\rangle_{{\rm max}}$.

\section{V. Conclusions}

We studied the statistics of quantum-heat in a quantum system subjected to a sequence of projective measurements of a generic observable. We have investigated the case when the time interval between consecutive measurements is a random variable. Previous works \cite{Campisi2011PRE,CampisiNJP2014,Campisi2011PRE,CampisiNJP2014} considered predetermined waiting times, showing that the quantum-heat obeys a integral fluctuation theorem reading like the Jarzynski equality where quantum-heat replaces the work. Here, we have shown that this continues to hold also when the waiting times are random. This result can be understood by noticing that the corresponding quantum dynamics is unital.

We have illustrated the theory for a two-level system, for which we have provided the explicit expressions of the characteristic function of quantum-heat. We have investigated both the annealed and the quenched distributions of waiting times. Interestingly, in the annealed case the quantum heat transferred by the two-level system to the heat bath is larger than the heat transferred in the quenched noise. Moreover, a larger transfer of quantum-heat can be observed in the stochastic case with respect to considering a sequence of measurements with fixed waiting times when the range of values of $\langle\tau\rangle$ is smaller than the one related to the energy splitting $\Delta E$, under the condition that $\langle\tau\rangle = \overline{\tau}$ (and therefore $\mathcal{T}$) is fixed. The latter condition can be referred to that of the stochastic quantum Zeno regime \cite{Gherardini2016NJP,Gherardini2017QSc,GherardiniPhDThesis}. Accordingly, our result reflects the intuition that a larger amount of noise in the waiting times between consecutive measurements of a given protocol is accompanied by a larger quantum-heat transfer. Finally, we have found the existence of a discontinuity in the characteristic function $G(u)$ when the protocol relies on the application of a large enough number of measurements of the Hamiltonian, i.e. $M\rightarrow\infty$ and $|a|^2\rightarrow 0,1$. This means that, in this limit, even an infinitesimal amount of noise in the measurement process will result into a finite amount of quantum-heat.

\section{Acknowledgments}

The authors gratefully acknowledge Giacomo Gori for fruitful discussions. S.G., M.M. and F.C. were financially supported from the Fondazione CR Firenze through the project Q-BIOSCAN. S.G. and L.B. also acknowledge the Scuola Internazionale Superiore di Studi Avanzati (SISSA) in Trieste for hospitality during the completion of this work.

\appendix

\section{Appendix}

\section{A1. Derivation of the characteristic function $G(u)$}

Here we derive the expression for the characteristic function
\begin{equation}\label{G_u_app}
G(u) = \int P(Q_{q})e^{iuQ_{q}}dQ_{q}
\end{equation}
by taking into account, respectively, quenched and annealed disorder for the waiting times between measurements. In Eq. (\ref{G_u_app}) the quantum-heat probability distribution is defined as
\begin{equation}
P(Q_{q}) = \sum_{n,m}\delta(Q_{q} - E_{m} + E_{n})p_{m|n}~p_{n},
\end{equation}
where $p_{m|n}$ is the transition probability to get the final energy $E_{m}$ conditioned to have measured $E_{n}$ after the first energy measurement.

\subsubsection{Quenched disorder}

Inserting the expression of the joint distribution
\begin{equation}
p(\vec{\tau}) = p(\tau_1) \prod_{i=2}^M \delta(\tau_i-\tau_1)
\end{equation}
into Eq. (\ref{conditional_prob_an}) in the main text, we obtain for the transition probability $p_{m|n}$ the expression
\begin{equation}
p_{m|n} = \sum_{\vec{k}}\int d\tau p(\tau){\rm Tr}\left[\Pi_{m}\mathcal{V}(\vec{k},\tau)\Pi_{n}\mathcal{V}^{\dagger}(\vec{k},\tau)\Pi_{m}\right],
\end{equation}
where $\mathcal{V}(\vec{k},\tau) = \Pi_{k_{M}}\mathcal{U}(\tau)\cdots\Pi_{k_1}\mathcal{U}(\tau)$. Accordingly, the corresponding quantum-heat probability distribution is equal to
\begin{equation}
P(Q_{q}) = \int \sum_{n,\vec{k},m}{\rm Tr}\left[\Pi_{m}\mathcal{V}(\vec{k},\tau)\Pi_{n}\mathcal{V}^{\dagger}(\vec{k},\tau)\Pi_{m}\right]p_{n}p(\tau)d\tau,
\end{equation}
so that the characteristic function $G(u)$ reads
\begin{eqnarray}
G(u) &=& \int\sum_{n,\vec{k},m}\bra{E_{m}}\mathcal{V}(\vec{k},\tau)\ket{E_{n}}\bra{E_{n}}\mathcal{V}^{\dagger}(\vec{k},\tau)\ket{E_{m}}\nonumber \\
&\cdot&e^{iu(E_{m} - E_{n})}p_{n}p(\tau)d\tau,
\end{eqnarray}
where we used the relation
\begin{equation}
{\rm Tr}\left[\Pi_{m}\mathcal{V}\Pi_{n}\mathcal{V}^{\dagger}\Pi_{m}\right]=\bra{E_{m}}\mathcal{V}\ket{E_{n}}\bra{E_{n}}\mathcal{V}^{\dagger}\ket{E_{m}}.
\end{equation}
Finally, using
\begin{equation}\label{eq:rel_eigenvalue_eq}
\begin{cases}
e^{iuE_{m}}\ket{E_{m}} = e^{iuH}\ket{E_{m}} \\
\bra{E_{n}}e^{-iuE_{n}} = \bra{E_{n}}e^{-iuH}
\end{cases},
\end{equation}
we obtain
\begin{eqnarray}
G(u) &=& \sum_{\vec{k}}\int\sum_{n,m}
\bra{E_{m}}\mathcal{V}\ket{E_{n}}\bra{E_{n}}\rho_{0}\ket{E_{n}}\nonumber \\
&\cdot&\bra{E_{n}}e^{-iuH}\mathcal{V}^{\dagger}e^{iuH}\ket{E_{m}}p(\tau)d\tau\nonumber \\
&=& \sum_{\vec{k}}\int{\rm Tr}\left[\mathcal{V}e^{-iuH}\rho_{0}\mathcal{V}^{\dagger}e^{iuH}\right]p(\tau)d\tau,
\end{eqnarray}
i.e. Eq. (\ref{G_u_finale}) in the main text for the quenched disorder case.

\subsubsection{Annealed disorder}

In case the stochasticity between consecutive projective measurements is distributed as an annealed disorder,
the joint distribution of the waiting times is
\begin{equation}
p(\vec{\tau}) = \prod_{j=1}^{M}p(\tau_{j}),
\end{equation}
so that the transition probability $p_{m|n}$ is given by
\begin{equation}
p_{m|n} = \sum_{\vec{k}}\int d^{M}\vec{\tau}p(\vec{\tau}){\rm Tr}\left[\Pi_{m}\mathcal{V}(\vec{k},\vec{\tau})\Pi_{n}\mathcal{V}^{\dagger}(\vec{k},\vec{\tau})\Pi_{m}\right].
\end{equation}
The latter corresponds to a multiple integral defined over the waiting times $\vec{\tau}$, where $\mathcal{V}(\vec{k},\vec{\tau}) = \Pi_{k_{M}}\mathcal{U}(\tau_{M})\cdots\Pi_{k_1}\mathcal{U}(\tau_{1})$. As a result, the quantum-heat probability distribution $P(Q_{q})$ and the corresponding characteristic function $G(u)$ can be written, respectively, as

\begin{small}
\begin{equation}
P(Q_{q}) = \sum_{\vec{k}}\int\sum_{n,m}{\rm Tr}\left[\Pi_{m}\mathcal{V}(\vec{k},\vec{\tau})\Pi_{n}\mathcal{V}^{\dagger}(\vec{k},\vec{\tau})\Pi_{m}\right]p_{n}p(\vec{\tau})d^{M}\vec{\tau}
\end{equation}
\end{small}

and
\begin{eqnarray}
G(u) &=& \sum_{\vec{k}}\int\sum_{n,m}
\bra{E_{m}}\mathcal{V}\ket{E_{n}}\bra{E_{n}}\rho_{0}\ket{E_{n}}\nonumber \\
&\cdot&\bra{E_{n}}e^{-iuH}\mathcal{V}^{\dagger}e^{iuH}\ket{E_{m}}p(\vec{\tau})d^{M}\vec{\tau}.
\end{eqnarray}
Accordingly, by using again the relations of Eq.~(\ref{eq:rel_eigenvalue_eq}), we can derive the expression of $G(u)$ as
\begin{equation}\label{G_u_finale_appendice_bis}
G(u) = \overline{\left\langle {\rm Tr}\left[e^{iuH}\mathcal{V}(\vec{k},\vec{\tau})e^{-iuH}\rho_{0}\mathcal{V}^{\dagger}(\vec{k},\vec{\tau})\right] \right\rangle},
\end{equation}
i.e. Eq. (\ref{G_u_finale}) in the main text, where the angular bracket denote quantum-mechanical expectation, while the overline stands for the average over the noise realizations.

%

\section{A2. Analytical $G(u)$ for a two-level system}

\subsubsection{Fixed waiting times sequence}

Let us consider a sequence of projective measurements applied to a $n-$level quantum system at fixed waiting times.
We denote with $\overline{\tau}$ the (fixed) time between consecutive measurements. The characteristic function of the quantum-heat is given by Eq.~(\ref{eq:G(u)_regular}), which can be rewritten as:
\begin{equation}
G(u) = f(u)L^{M-1}g(u).
\end{equation}
For a two-level system an explicit expression for $G(u)$ can be derived. To this end, we assume, without loss of generality, that the system energy values $E_{\pm}$ are equal to $\pm E$ and, then, we make use of the energy eigenvalue equation, i.e. $H|E_{\pm}\rangle = E_{\pm}|E_{\pm}\rangle$, so as to obtain
\begin{equation}\label{B2}
f(u)^{T} = \begin{pmatrix}
\langle\alpha_{1}|e^{iuH}|\alpha_{1}\rangle \\
\langle\alpha_{2}|e^{iuH}|\alpha_{2}\rangle
\end{pmatrix} =
\begin{pmatrix}
|a|^{2}e^{iuE} + |b|^{2}e^{-iuE} \\
|a|^{2}e^{-iuE} + |b|^{2}e^{iuE}
\end{pmatrix},
\end{equation}
where $\{|\alpha_{j}\rangle\}$, $j = 1,2$, is the basis, defining the projective measurements of the protocol.
The elements of the basis $\{|\alpha_{j}\rangle\}$ are chosen as linear combinations of the energy eigenstates $|E_{\pm}\rangle$, see Eq.~(\ref{eq:abasis}). The transition matrix $L$ turns out to be
\begin{eqnarray}\label{B3}
L &=& \begin{pmatrix}
\left||a|^{2}e^{-iEt} + |b|^{2}e^{iEt} \right|^{2} &
\left|a^{\ast}be^{-iEt} - ab^{\ast}e^{iEt} \right|^{2} \\ \left|b^{\ast}ae^{-iEt} - ba^{\ast}e^{iEt} \right|^{2} & \left||b|^{2}e^{-iEt} + |a|^{2}e^{iEt} \right|^{2}
\end{pmatrix}\nonumber \\
&=&\begin{pmatrix}
1-\overline{\nu} & \overline{\nu} \\ \overline{\nu} & 1-\overline{\nu}
\end{pmatrix},
\end{eqnarray}
where
\begin{equation}\label{eq:nu}
\overline{\nu} \equiv 2|a|^{2}|b|^{2}\sin^{2}(\overline{\tau}\Delta E),
\end{equation}
and $\Delta E \equiv(E_{+} - E_{-}) = 2E$. By using the decomposition of the initial density matrix $\rho_{0}$ in the energy basis, Eq.~(\ref{eq:rho_0_2_level_system})), and again Eq.~(\ref{eq:abasis}), it holds that
\begin{equation}\label{eq:g(u)}
g(u) = \begin{pmatrix}
\langle\alpha_{1}|e^{-iuH}\rho_{0}|\alpha_{1}\rangle \\
\langle\alpha_{2}|e^{-iuH}\rho_{0}|\alpha_{2}\rangle
\end{pmatrix}
= \begin{pmatrix}
|a|^{2}c_{1}e^{-iuE}+|b|^{2}c_{2}e^{iuE}\\
|a|^{2}c_{2}e^{iuE} + |b|^{2}c_{1}e^{-iuE}
\end{pmatrix}.
\end{equation}
In conclusion, the explicit dependence of $G(u)$ from the set of parameters $(a,b,c_{1},c_{2},\overline{\tau})$ is given by the following equation:
\begin{eqnarray}\label{eq:G_ordered_2LS}
G(u)&=&\begin{pmatrix}
|a|^{2}e^{iuE}+|b|^{2}e^{-iuE}\\
|a|^{2}e^{-iuE} + |b|^{2}e^{iuE}
\end{pmatrix}^{T}\begin{pmatrix}
1-\overline{\nu} & \overline{\nu} \\ \overline{\nu} & 1-\overline{\nu}
\end{pmatrix}^{M-1}\nonumber \\
&\cdot&\begin{pmatrix}
|a|^{2}c_{1}e^{-iuE}+|b|^{2}c_{2}e^{iuE}\\
|a|^{2}c_{2}e^{iuE} + |b|^{2}c_{1}e^{-iuE}
\end{pmatrix}.
\end{eqnarray}
It is worth noting that the characteristic function $G(u)$ admits a discontinuity point in correspondence of $|a|^2\rightarrow 0,1$ and $M\rightarrow\infty$. In particular, when $|a|^2\rightarrow 0,1$ and a finite number $M$ of measurements is performed, $G(u)$ is identically equal to $1$. Conversely, under the asymptotic limit $M\rightarrow\infty$, the characteristic function does not longer depend on $a$ and it equals to
\begin{equation}
G(u) = \frac{(1+e^{2iuE})}{2} - c_{1}\sinh(2iuE),
\label{eq:B9}
\end{equation}
so that
$G(i\beta) = (1+e^{-2\beta E})/2 + c_{1}\sinh(2\beta E)$. The transition matrix $L$ admits as eigenvalues the values $1$ and $(1 - 2\overline{\nu})<1$ .Thus, after the eigendecomposition of the transition matrix, for $M\rightarrow\infty$
one finds that
\begin{equation}
L^{M-1} \longrightarrow V\begin{pmatrix} 0 & 0 \\ 0 & 1 \end{pmatrix}V^{T} = \begin{pmatrix} \frac{1}{2} & \frac{1}{2} \\ \frac{1}{2} & \frac{1}{2} \end{pmatrix},
\end{equation}
with
\begin{equation}
V = \begin{pmatrix}
     -\frac{1}{\sqrt{2}} & \frac{1}{\sqrt{2}} \\
     \frac{1}{\sqrt{2}} & \frac{1}{\sqrt{2}}
   \end{pmatrix}.
\end{equation}

\subsubsection{Stochastic waiting times sequence}

Here we take into account a sequence of projective measurements with stochastic waiting times $\tau_{k}$, $k = 1,\ldots,M$, sampled by a bimodal probability density function $p(\tau)$.

The explicit expression of the characteristic function in the presence of quenched disorder can be derived from Eqs. (\ref{G_u_finale}) and (\ref{eq:G_ordered_2LS}). We obtain

\begin{small}
\begin{eqnarray}\label{B7}
G(u)&=&\sum_{j = 1}^{d_{\tau}}
\begin{pmatrix}
|a|^{2}e^{iuE}+|b|^{2}e^{-iuE}\\
|a|^{2}e^{-iuE} + |b|^{2}e^{iuE}
\end{pmatrix}^T
\begin{pmatrix}
1-\nu_{j} & \nu_{j} \\
\nu_{j} & 1-\nu_{j}
\end{pmatrix}^{M-1}\nonumber \\
&\cdot&\begin{pmatrix}
|a|^{2}c_{1}e^{-iuE}+|b|^{2}c_{2}e^{iuE}\\
|a|^{2}c_{2}e^{iuE} + |b|^{2}c_{1}e^{-iuE}
\end{pmatrix}p_j,
\end{eqnarray}
\end{small}

where
\begin{equation}\label{eq:eta}
\nu_{j}\equiv\nu(\tau^{(j)})=
2|a|^{2}|b|^{2}\sin^{2}(2\tau^{(j)}E),
\end{equation}
and $d_{\tau} = 2$ is the number of values that can be assumed by $\tau$.
\begin{figure}[t]
\includegraphics[scale=0.48]{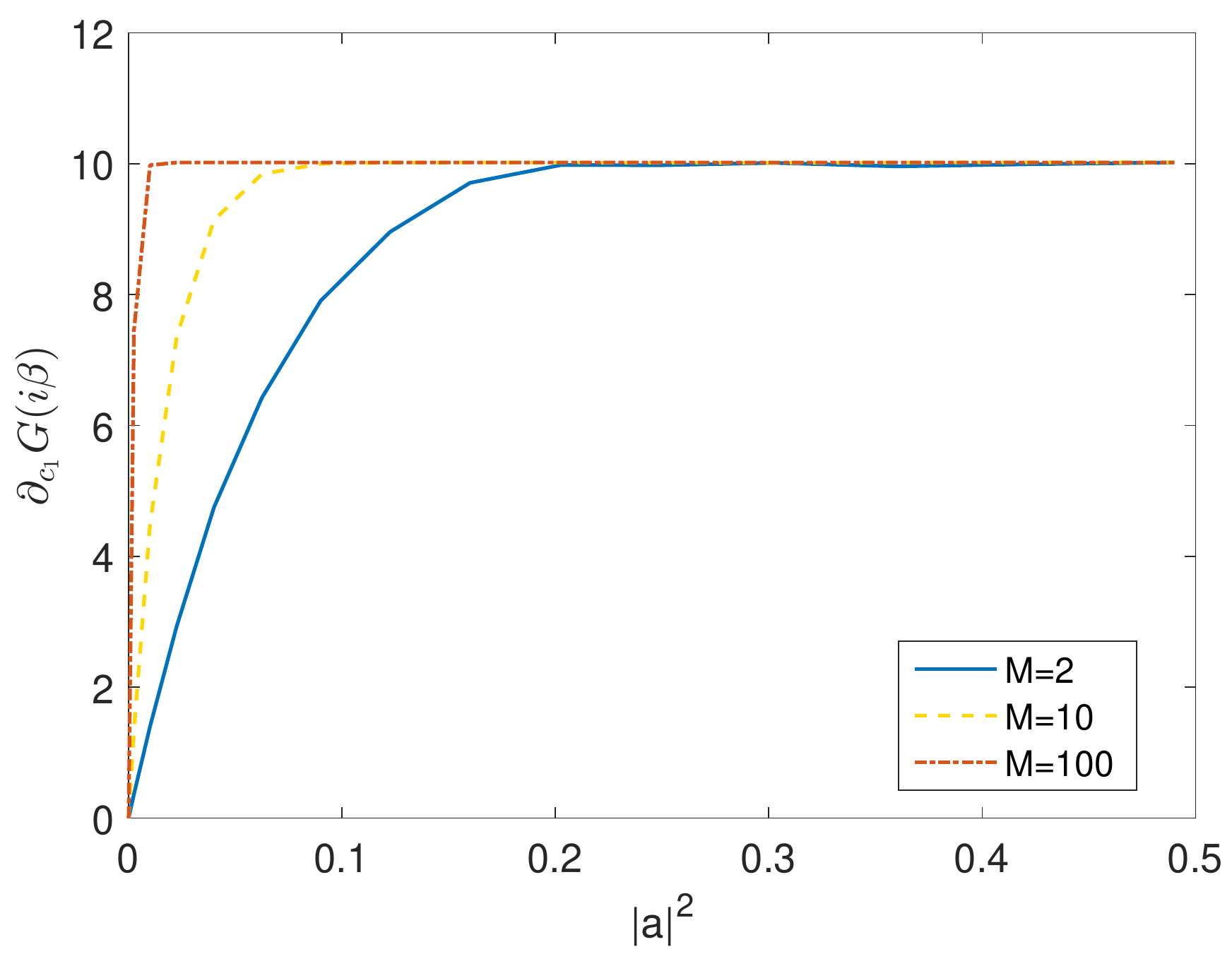}
\caption{Convergence of $\partial_{c_1}G(i\beta)$ (slope of $G(i\beta)$ w.r.t. $c_1$) as a function of $|a|^2$ ($|a|^2\in[0,0.5]$) to the asymptotic limit of $M\rightarrow\infty$ when the time intervals between measurements are distributed as quenched disorder. We have considered $M = 2$ (solid blue), $M = 10$ (dashed yellow) and $M = 100$ (dash-dotted orange), with $E_{\pm} = \pm 1$, $\tau^{(1)} = 0.01$, $\tau^{(2)} = 3$ and $p_{1} = 0.3$.}
\label{fig:fig5}
\end{figure}

As discussed in the main text, also in this case the characteristic function admits a discontinuity point in correspondence of $|a|^2\rightarrow 0,1$ and $M\rightarrow\infty$. As before, when $|a|^2\rightarrow 0,1$ and a finite number $M$ of measurements is performed, $G(u)$ is identically equal to $1$; while for $M\rightarrow\infty$ the characteristic function does not longer depend on $a$ and it equals again to
\begin{equation}
G(u) = \frac{(1+e^{2iuE})}{2} - c_{1}\sinh(2iuE),
\label{G_u_qu}
\end{equation}
as we obtained in the non-stochastic case. Indeed, the transition matrix $L\left(\tau^{(j)}\right)$ admits as eigenvalues the values $1$ and $(1 - 2\nu_{j})<1$, so that for $M\rightarrow\infty$ one has
\begin{equation}
L\left(\tau^{(j)}\right)^{M-1} \longrightarrow \begin{pmatrix} \frac{1}{2} & \frac{1}{2} \\ \frac{1}{2} & \frac{1}{2} \end{pmatrix}~~\text{with}~~j=1,\ldots,d_{\tau}.
\end{equation}
For the sake of clarity, let us observe the results shown in Fig.~\ref{fig:fig5}, where the slope of $G(i\beta)$ w.r.t. $c_1$, i.e. $\partial_{c_1}G(i\beta)$, changes for different values of $|a|^2\in[0,1/2]$ and increasing values of $M$.

Finally, we repeat the latter derivation when the stochasticity between measurements is distributed as annealed disorder. In this regard, the characteristic function can be written as
\begin{eqnarray}
G(u) &=& \sum_{k = 0}^{M-1}\binom{M-1}{k}f(u)L(\tau^{(1)})^{k}L(\tau^{(2)})^{M-k-1}\nonumber \\
&\cdot&g(u)p_1^{k}p_2^{M-k-1}.
\label{eq:B12}
\end{eqnarray}
By substituting the expressions of $f(u)$, $L$ and $g(u)$ as given in Eqs.~(\ref{B2}),~(\ref{B3}),~(\ref{eq:nu}) and (\ref{eq:g(u)}), we obtain the following relation:

\begin{small}
\begin{eqnarray}
G(u)&=&\sum_{k = 0}^{M-1}\binom{M-1}{k}\begin{pmatrix}
|a|^2e^{iuE}+|b|^2e^{-iuE}\\
|a|^2e^{-iuE}+|b|^2e^{iuE}
\end{pmatrix}^T \nonumber\\
&\cdot&\begin{pmatrix}
1-\nu_{1} & \nu_{1} \\ \nu_{1} & 1-\nu_{1}
\end{pmatrix}^{k} \begin{pmatrix}
1-\nu_{2} & \nu_{2} \\ \nu_{2} & 1-\nu_{2}
\end{pmatrix}^{M-k-1} \nonumber\\
&\cdot& \begin{pmatrix}
|a|^2c_{1}e^{-iuE}+|b|^2c_{2}e^{iuE}\\
|a|^2c_{2}e^{iuE}+|b|^2c_{1}e^{-iuE}
\end{pmatrix} p_1^{k}p_2^{M-k-1},
\end{eqnarray}
\end{small}

As for the other cases, we find the same discontinuity in $G(u)$ in the limits of $|a|^2\rightarrow 0,1$ and $M\rightarrow\infty$. Quite surprisingly, the discontinuity is exactly the same for both types of disorder. To observe this, let us take Eq.~(\ref{eq:B12}) with $a \neq 0$ and use the binomial theorem,
which states that $\displaystyle{(x+y)^{n}=\sum_{k=0}^{n} \binom{n}{k}x^{n-k}y^{k}}$ with $x$, $y$ arbitrary real variables. As a result, we obtain
\begin{equation}
G(u) = f(u)\left(L(\tau^{(1)})p_1+L(\tau^{(2)})p_2\right)^{M-1}g(u).
\end{equation}
By introducing the quantity $\zeta\equiv\nu_1p_1+\nu_2p_2$, the weighted sum [w.r.t. $p(\tau)$] of the transition matrices $L(\tau^{(1)})$ and $L(\tau^{(2)})$ can be simplified as
\begin{equation}
\left(L(\tau^{(1)})p_1+L(\tau^{(2)})p_2\right)=
\begin{pmatrix}
1-\zeta & \zeta \\ \zeta & 1-\zeta
\end{pmatrix},
\end{equation}
which admits eigenvalues 1 and $(1-2\zeta)\leq 1$. Thus, by performing the limit $M\rightarrow\infty$, the weighted sum of the transition matrices tends to a projector,
so that $G(u)$ is effectively given by Eq.~(\ref{G_u_qu}).

\subsubsection{$n-$th order derivative of $G(u)$}

Analytical expression for $\partial^{n}_{u}G(u)$ allows to derive all the statistical moments of the quantum-heat, and, consequently, its mean value $\langle Q_{q}\rangle$. In particular, the $n-$th order derivative of the quantum-heat characteristic function, when a protocol of projective measurements at fixed waiting times is considered, is
\begin{equation}\label{eq:partial_derivative}
\partial^{n}_{u}G(u)=
\sum_{k=0}^{n} A^k(u)^{T}\cdot\begin{pmatrix}
1-\overline{\nu} & \overline{\nu} \\ \overline{\nu} & 1-\overline{\nu}
\end{pmatrix}^{M-1} \cdot B^{n-k}(u),
\end{equation}
where
\begin{equation}
A^l(u) \equiv
(i)^{l}
\begin{pmatrix}
\bra{\alpha_{1}}H^{l}e^{iuH}
\ket{\alpha_{1}} \\
\bra{\alpha_{2}}H^{l}e^{iuH}
\ket{\alpha_{2}}
\end{pmatrix}
\end{equation}
and
\begin{equation}
B^l(u)\equiv
(-i)^{l}
\begin{pmatrix}
\bra{\alpha_{1}}H^{l}e^{-iuH}\rho_{0}
\ket{\alpha_{1}} \\
\bra{\alpha_{2}}H^{l}e^{-iuH}\rho_{0}
\ket{\alpha_{2}}
\end{pmatrix}.
\end{equation}
In the quenched disorder case $\partial^{n}_{u}G(u)$ reads
\begin{eqnarray}\label{eq:partial_derivative_quench}
\partial^{n}_{u}G(u)&=&\displaystyle{\sum_{j = 1}^{d_{\tau}}
\sum_{k=0}^{n}A^k(u)^{T}}\cdot\begin{pmatrix}
1-\nu(\tau^{(j)}) & \nu(\tau^{(j)})\\
\nu(\tau^{(j)}) & 1-\nu(\tau^{(j)})
\end{pmatrix}^{M-1}\nonumber \\
&\cdot& B^{n-k}(u)p_{j},
\end{eqnarray}
while in the annealed case
\begin{eqnarray}\label{eq:partial_derivative_annealed}
&&\partial^{n}_{u}G(u) = \displaystyle{\sum_{k = 0}^{M-1}\sum_{l=0}^{n}A^{l}(u)^{T}}
\cdot\begin{pmatrix}
1-\nu(\tau^{(1)}) & \nu(\tau^{(1)})\\
\nu(\tau^{(1)}) & 1-\nu(\tau^{(1)})
\end{pmatrix}^{k}\nonumber \\
&&\cdot\begin{pmatrix}
1-\nu(\tau^{(2)}) & \nu(\tau^{(2)})\\
\nu(\tau^{(2)}) & 1-\nu(\tau^{(2)})
\end{pmatrix}^{M-k-1}\cdot B^{n-l}(u)p_{1}^{k}p_{2}^{M-k-1}.\nonumber \\
&&
\end{eqnarray}


\begin{thebibliography}{99}

\bibitem{Esposito2009}
M. Esposito, U. Harbola, and S. Mukamel. Nonequilibrium fluctuations, fluctuation theorems, and counting statistics in quantum systems. Rev. Mod. Phys. {\bf 81}, 1665 (2009).

\bibitem{Campisi2011}
M. Campisi, P. Hanggi, and P. Talkner. {\it Colloquium}: Quantum fluctuations relations: Foundations and applications. Rev. Mod. Phys. {\bf 83}, 1653 (2011).

\bibitem{Scully03Science299}
M.O. Scully, M.S. Zubairy, G.S. Agarwal, and H. Walther. Extracting work from a single heat bath via vanishing quantum coherence. Science {\bf 299}, 862 (2003).

\bibitem{Kosloff14ARPC65}
R. Kosloff and A. Levy. Quantum Heat Engines and Refrigerators: Continuous Devices. Annual Review of Physical Chemistry {\bf 65}, 365 (2014).

\bibitem{Uzdin15PRX5}
R. Uzdin, A. Levy, and R. Kosloff. Equivalence of Quantum Heat Machines, and Quantum-Thermodynamic Signatures. Phys. Rev. X {\bf 5}, 031044 (2015).

\bibitem{Campisi15NJP17}
M. Campisi, J. Pekola, and R. Fazio. Nonequilibrium fluctuations in quantum heat engines: theory, example, and possible solid state experiments. New J. Phys. {\bf 17}, 035012 (2015).

\bibitem{Solinas2015}
P. Solinas and S. Gasparinetti. Full distribution of work done on a quantum system for arbitrary initial states. Phys. Rev. E {\bf 92}, 042150 (2015).

\bibitem{Campisi2010PRL}
M. Campisi, P. Talkner, and P. H\"{a}nggi. Fluctuation Theorems for Continuously Monitored Quantum Fluxes. Phys. Rev. Lett. {\bf 105}, 140601 (2010).

\bibitem{Campisi2011PRE}
M. Campisi, P. Talkner, and P. H\"{a}nggi. Influence of measurements on the statistics of work performed on a quantum system. Phys. Rev. E {\bf 83}, 041114 (2011).

\bibitem{Yi2013}
J. Yi and Y.W. Kim. Nonequilibirum work and entropy production by quantum projective measurements. Phys. Rev. E {\bf 88}, 032105 (2013).

\bibitem{WatanabePRE2014}
G. Watanabe, B. Prasanna Venkatesh, P. Talkner, M. Campisi, and P. H\"{a}nggi. Quantum fluctuation theorems and generalized measurements during the force protocol. Phys. Rev. E {\bf 89}, 032114 (2014).

\bibitem{Talkner16PRE93}
P. Talkner and P. H\"{a}nggi. Aspects of quantum work. Phys. Rev. E {\bf 93}, 022131 (2016).

\bibitem{Yuanjian16PRE94}
Y. Zheng, P. Hänggi, and D. Poletti. Occurrence of discontinuities in the performance of finite-time quantum Otto cycles. Phys. Rev. E {\bf 94}, 012137 (2016).

\bibitem{Elouard2016}
C. Elouard, D.A. Herrera-Mart\`i, M. Clusel and A. Auff\'eves. The role of quantum measurement in stochastic thermodynamics, NJP Quantum Info. {\bf 3}, 9 (2017).

\bibitem{Gherardini2016NJP}
S. Gherardini, S. Gupta, F.S. Cataliotti, A. Smerzi, F. Caruso, and S. Ruffo. Stochastic quantum Zeno by large deviation theory. New J. Phys. {\bf 18}, 013048 (2016).

\bibitem{Mueller2017ADP}
M.M. M\"{u}ller, S. Gherardini, and F. Caruso. Quantum Zeno dynamics through stochastic protocols. Annalen der Physik {\bf 529}, 1600206 (2017).

\bibitem{GherardiniPhDThesis}
S. Gherardini. Noise as a resource. PhD Thesis (University of Florence, Italy, 2018).

\bibitem{SchaferZeno}
F. Sch\"afer, I. Herrera, S. Cherukattil, C. Lovecchio, F.S. Cataliotti, F. Caruso, and A. Smerzi. Experimental realization of quantum {Z}eno dynamics. Nat. Commun. {\bf 5}, 4194 (2014).

\bibitem{Gherardini2017QSc}
S. Gherardini, C. Lovecchio, M.M. M\"{u}ller, P. Lombardi, F. Caruso, and F.S. Cataliotti. Ergodicity in randomly perturbed quantum systems, Quantum Sci. Technol. {\bf 2}, 015007 (2017).

\bibitem{Albash13PRE88}
T. Albash, D.A. Lidar, M. Marvian, and P. Zanardi. Fluctuation theorems for quantum process. Phys. Rev. A {\bf 88}, 023146 (2013).

\bibitem{Rastegin13JSTAT13}
A.E. Rastegin. Non-equilibrium equalities with unital quantum channels. J. Stat. Mech. P06016 (2013).

\bibitem{Kafri2012}
D. Kafri and S. Deffner. Holevo's bound from a general quantum fluctuation theorem. Phys. Rev. A {\bf 86}, 044302 (2012).

\bibitem{Campisi17NJP19}
M. Campisi, J. Pekola, and R. Fazio. Feedback-controlled heat transport in quantum devices: theory and solid-state experimental proposal. New J. Phys. {\bf 19}, 053027 (2017).

\bibitem{Mezard1987}
M. Mezard, G. Parisi, and M.A. Virasoro. Spin glass theory and beyond (Singapore, World Scientific, 1987).

\bibitem{Sakurai1994}
J.J. Sakurai. Modern Quantum Mechanics (2nd edition, 1994).

\bibitem{Dorner2013}
R. Dorner, S.R. Clark, L. Heaney, R. Fazio, J. Goold, and V. Vedral. Extracting Work Statistics and Fluctuation Theorems by Single-Qubit Interferometry. Phys. Rev. Lett. {\bf 110}, 230601 (2013).

\bibitem{MazzolaPRL2013}
L. Mazzola, G. De Chiara, and M. Paternostro. Measuring the Characteristic Function of the Work Distribution. Phys. Rev. Lett. {\bf 110}, 230602 (2013).

\bibitem{CampisiNJP2014}
M. Campisi, R. Blattmann, S. Kohler, D. Zueco, and P. H\"{a}nggi. Employing circuit QED to measure non-equilibrium work fluctuations. New J. Phys. {\bf 15}, 105028 (2014).

\bibitem{GherardiniEntropy}
S. Gherardini, M.M. Mueller, A. Trombettoni, S. Ruffo, and F. Caruso. Reconstruction of the stochastic quantum entropy production to probe irreversibility and correlations. \verb|arXiv:1706.02193|

\bibitem{Q_Zeno_noise_detection}
M.M. M\"{u}ller, S. Gherardini, and F. Caruso. Stochastic quantum {Z}eno-based detection of noise correlations. Sci. Rep. {\bf 6}, 38650 (2016).

\end{thebibliography}
\end{document}